\documentclass[%
 reprint,
superscriptaddress,
 amsmath,amssymb,
 aps,
pre,
]{revtex4-2}

\usepackage{graphicx}
\usepackage{dcolumn}
\usepackage{bm}
\usepackage{xcolor}

\begin{document}

\title{Loading-dependent microscale measures control bulk properties in granular material: an experimental test of the Stress-Force-Fabric relation}
\author{Carmen L. Lee}
\affiliation{Department of Physics, North Carolina State University, Raleigh, North Carolina, 27695, USA}

\author{Ephraim Bililign}%
\affiliation{Department of Physics, North Carolina State University, Raleigh, North Carolina, 27695, USA}
\affiliation{James Franck Institute and Department of Physics, University of Chicago, Chicago, IL, USA}

\author{Emilien Az\'ema}
\affiliation{LMGC, Universit\'e de Montpellier, CNRS, Montpellier, France}
\affiliation{Department of Civil, Geological, and Mining Engineering, Polytechnique Montréal, Montréal, Canada.}
\affiliation{Institut Universitaire de France (IUF), Paris, France}

\author{Karen E. Daniels}
\affiliation{Department of Physics, North Carolina State University, Raleigh, North Carolina, 27695, USA}

\begin{abstract}
    The bulk behaviour of granular materials is tied to its mesoscale and particle-scale features: strength properties arise from the buildup of various anisotropic structures at the particle-scale induced by grain connectivity (fabric), force transmission, and frictional mobilization. More fundamentally, these anisotropic structures work collectively to define features like the bulk friction coefficient and the stress tensor at the macroscale and can be explained by the Stress-Force-Fabric (SFF) relationship stemming from the microscale. Although the SFF relation has been extensively verified by discrete numerical simulations, a laboratory realization has remained elusive due to the challenge of measuring both normal and frictional contact forces. In this study, we analyze experiments performed on a photoelastic granular system under four different loading conditions: uniaxial compression, isotropic compression,  pure shear, and annular shear. During these experiments, we record particle locations, contacts, and normal and frictional forces to measure the particle-scale response to progressing strain. We track microscale measures like the packing fraction, average coordination number and average normal force along with anisotropic distributions of contacts and forces. We match the particle-scale anisotropy to the bulk using the SFF relation, which is founded on two key principles, a Stress Rule to describe the stress tensor and a Sum Rule to describe the bulk friction coefficient; we find that the Sum and Stress Rules accurately describe bulk measurements. Additionally, we test the assumption that fabric and forces transmit load equally through our granular packings and show that this assumption is sufficient at large strain values, and can be applied to areas like rock mechanics, soft colloids, or cellular tissue where force information is inaccessible.
\end{abstract}
\maketitle

\section{Introduction}

When particulate materials like dry grains, dense colloidal suspensions, or cellular tissues are subjected to an external load, the bulk response is the result of intricate, cumulative interactions between the constituent parts. Consequently, bulk properties like shear modulus, elastic modulus, and yield strength are not well-defined material properties, but instead arise from the particular interior structure of the assembly of particles~\cite{liu_jamming_2010, vanHecke_jamming_2009}. As such, particle-scale information is required to accurately predict bulk properties; yet, this information will evolve over time and it is not easily accessible in most non-idealized scenarios.

The importance of the internal structure is seen in simulated frictionless systems of soft grains, measurements on a single configuration exhibit an anisotropic shear modulus that strongly depends on the shear angle, and different packings of the same particles exhibit different shear moduli under identical loading conditions~\cite{dagois-bohy_soft-sphere_2012}. This indicates that fabric, or inter-particle contacts are important determiners of bulk modulus. As well, previous works have used changes in fabric to explain rheological properties through fabric-flow models~\cite{olsen_modeling_tensorial_2015, rojasparra_capturing_2019, sun_constitutive_model_2011}.

Anisotropic force chain structures appear when a granular material is under load, on average aligning with the principal stress direction~\cite{majmudar_contact_2005}. The presence of these mesoscale features have been used to address questions like the origin of jamming~\cite{cates_jamming_1998}, plastic deformations~\cite{meng_force_chain_2021} and the heterogeneous distribution of stress~\cite{hurley_quantifying_2016}. Recent studies have used force chains to study local stress distributions~\cite{wang_contact_2021} and emergent elasticity~\cite{nampoothiri_emergent_2020, nampoothiri_tensor_2022}. Not only does the loading affect the bulk properties, but it also has a lasting effect on the dynamics of the material. Granular materials in particular  store memory of loading through history dependence due to the absence of internal driving to restore the system to equilibrium~\cite{keim_mechanical_2022}. 
Thus, the bulk properties of a granular packing under load depends on the loading geometry and the emergence of anisotropic structures (force chains and contacts) during loading to ensure the mechanical strength of the assembly.

As such, we need a way to record the complete microstructural information using quantifiable, understandable parameters. The spatial organization of the particles and contacts, making up a granular texture (fabric), can take on highly variable morphologies in terms of particle connectivity, particle orientation (in the case of non-spherical grains), contact plane orientation and force transmission; these characteristics evolves with deformation.  
For example, a well-known, simple characteristic is the coordination number $z$ quantifying the mean number of contacts per particles. 
To provide a further level of detail about spatial heterogeneity, the anisotropy in the contacts can be described by probability distribution functions quantifying how the mean value of $z$ varies with the direction along which it is measured. Such measures can further be extend to other scalar properties, for instance the mean force, or the friction mobilization etc.

One well-explored method of quantifying textures using average particle-scale data is the Stress-Force-Fabric (SFF) relation originally developed by \citet{rothenburg_analytical_1989}. It provides a framework to understand how bulk properties in granular material (e.g. the stress tensor) arise from the anisotropy in contact vectors, vector forces and contact distribution in a granular assembly. \citet{li_stressforcefabric_2013} have since refined the original derivation using directional statistics to formalize these relationships. Over the decades, SFF has been used to study the effect of changing grain shapes~\cite{azemaNonlinearEffectsParticle2012, azemaPackingsIrregularPolyhedral2013, azemaStressstrainBehaviorGeometrical2010},  particle size distribution~\cite{voivretMultiscaleForceNetworks2009, azema_shear_2017, cantorMicrostructuralAnalysisSheared2020}, interparticle friction~\cite{bathurstObservationsStressforcefabricRelationships1990,binaree_combined_2020}, as well as to describe in detail the role of the strong/weak force networks in the stress transmission mechanisms~\cite{radjaiBimodalCharacterStress1998, azemaForceChainsContact2012, kruytWeakStrongContact2016}. 

However, because the SFF relation requires particle-scale information, it has yet to be fully applied to laboratory experiments. With improvements to imaging and computing power, we are now able to use photoelastic particles to test  the SFF relation in laboratory conditions and find that it is an accurate description for all of the conditions that we have applied it to. This provides a common framework with which to describe a diverse set of experimental results, and thereby make cross-comparisons.

In this work, we experimentally explore the role of loading geometry on the bulk stress tensor and bulk friction coefficient in granular material using assemblies of photoelastic disks. We present a laboratory realization of the SFF relation under four loading geometries: uniaxial compression, isotropic compression, pure shear, and annular shear. We map the evolution of the packing fraction, average coordination number and average force as a function of strain, as well as the evolution of the anisotropy in the distributions of the contact orientation and forces. The relative magnitudes of the anisotropic distributions are used to verify the SFF relation. We find that that the SFF relation accurately predicts the bulk stress tensor and bulk friction coefficient for all quasi-statically loaded histories using particle-scale anisotropies over multiple data sets.

\section{Methods}

\begin{figure*}
    \centering
    \includegraphics[width = \textwidth]{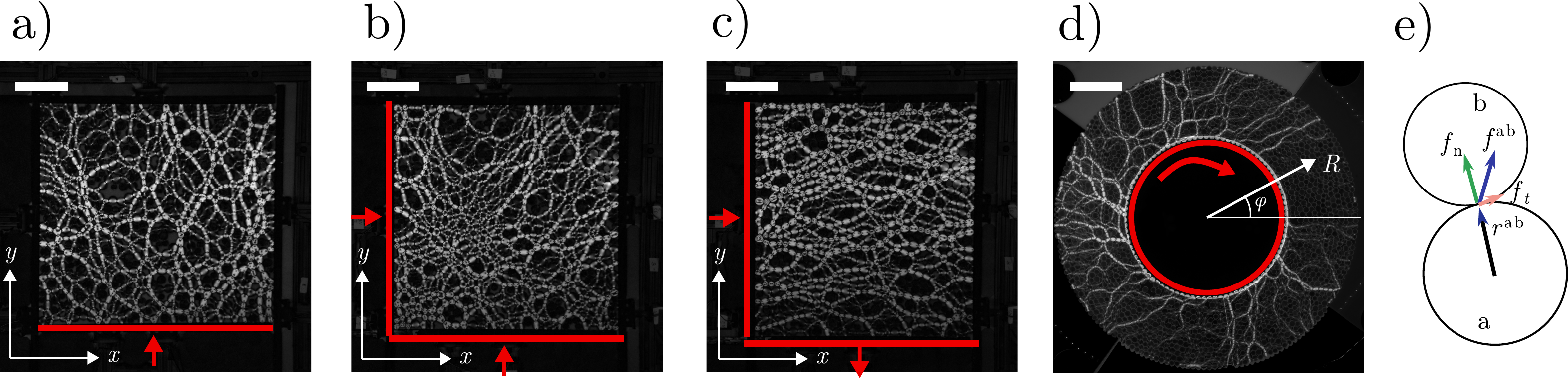}
    \caption{Images of the four loading geometries (a) Uniaxial compression (b) Isotropic compression (c) Pure shear (a)-(c) contain 890 bidisperse photoelastic particles that float on an airtable with no basal friction. (d) Annular shear loading of approximately 2000 bidisperse photoelastic particles on a supported surface. Loadings (a)-(c) follow a Cartesian coordinate system and (d) follows a polar coordinate system. Scale bars indicates 10 cm. (e) A schematic of two particles in contact showing the contact position vector $r^{ab}$, contact force vector $f^{ab}$, normal $f_n$ and tangential forces $f_t$. }
    \label{fig:loading}
\end{figure*}

We conduct experiments on granular packings subject to four different loading geometries. Three loadings are performed in a two-dimensional rectangular granular system composed of $N_p$ = 890 bidisperse (Vishay PhotoStress PSM-4, bulk modulus $E$= 4 MPa, density 1.06 g/cm$^3$) circular disks with an equal fraction of radii $R_s$ = 5.5 mm and $R_l$ = 7.7 mm ($R_l/R_s = 1.4$). The interparticle friction coefficient $\mu_p$ is approximately 0.3. A sub-fluidizing upflow of air passing through a porous polypropylene sheet floats the particles to eliminate basal friction~\cite{puckett_equilibrating_2013}. The packings are subjected to three quasi-static loading geometries as visualised in Fig.~\ref{fig:loading} where the photoelastic response of grains under (a) uniaxial compression, (b) isotropic compression and (c) pure (volume-preserving) shear. Further details regarding the experimental set up can be found in \citet{bililign_protocol_2019}; in the results presented here, the original images from these experiments have been reanalyzed to additionally detect small forces and contacts between particles. Each dataset consists of 30 statistically-independent packings loaded from their boundaries: after each loading cycle, the confining walls are retracted, the system dilates to a very loose packing, and a brief overhead flow of turbulent air rearranges the particles to produce a fresh  configuration. We previously found that an ensemble of size $\sim$30 is sufficient to produce well-defined means. 

To test the stress-force-fabric relation at larger strains, including reaching steady state, we performed a fourth type of loading in an annular geometry as shown  Fig.~\ref{fig:loading}(d). This quasi-two-dimensional annular shear cell, first described in~\cite{brzinski_sounds_2018} contains $N_p$ = 1810 bidisperse (also Vishay PhotoStress PSM-4, bulk modulus $E$= 4 MPa, density 1.06 g/cm$^3$, $\mu_p~\approx 0.3$) circular disks with an equal fraction of radii $R_s$ = 4.5 mm and $R_l$ = 5.5 mm ($R_l/R_s = 1.2$). The particles are dusted with baking powder as a dry lubricant to reduce basal friction between the particles and the supporting sheet. The inner and outer boundaries have particle-sized rough features to reduce slip between the boundaries and the grains. The inner boundary of the annulus is rotated at a constant rate of 1 rotation every 2.5 hours and the packing undergoes shear analogous to simple shear in an annular geometry. Using the annular strain rate $\Dot{\gamma} = \frac{1}{2} \left(\frac{\partial v}{\partial R} - \frac{v}{R}\right)$ as described by \citet{tang_nonlocal_2018}, and integrating over time with appropriate boundary conditions, we calculate the strain $\epsilon$.  Further details of the system and this new dataset can be found in the companion article~\citet{lee_preprint}, where we additionally investigate the long-term, transient dynamics. 

For all four loading geometries, we extract particle positions, contacts, and interparticle vector force data from the images of the packings. Images are analyzed using the open source PhotoElastic Grain Solver~\cite{kollmer_photo-elastic_nodate}. To perform these studies, we have modified the original code~\cite{lee_photo-elastic_nodate} to detect contacts with resolution of $\sim 5 \times 10^{-5}$ N, an order of magnitude lower than previously reported. Contacts were detected using a combination of the gradient of intensity squared method~\cite{abed_zadeh_enlightening_2019, daniels_photoelastic_2017} and peak finding for contacts below the gradient threshold (see SI).

\section{Macroscale metrics and the Stress-Force-Fabric relation}

From the particle-scale information, we calculate the stress tensor 
using the classical micromechanical definition~\cite{bagi_microstructural_1999, kruyt_micromechanical_2014}. The stress tensor is computed by summing the outer product between the contact force vector $\vec{f}$ and the contact position vector $\vec{r}$ between particles $a$ and $b$
\begin{equation}
    \sigma_{ij} = \frac{1}{S} \sum_{a \neq b} r^{ab}_i \, f^{ab}_j
    \label{eq:stresstensor}
\end{equation}
which is normalized by area of the packing chamber $S$. 

As illustrated in Fig.~\ref{fig:loading}(e), the force vector $\vec{f}$ points from the location of the contact in the direction of the force on the particle. The contact position vector $\vec{r}$ points from the center of the particle to the contact. The summation takes place over all particles $a$ that have $b$ neighbours and it is assumed that $f^{ab} = -f^{ba}$ due to force balance between particles. Cycled over each stress direction $i$ and $j$ ($x$ and $y$) or ($R$ and $\phi$ for the annular cell), this micromechanical model builds a two dimensional tensor equal to the bulk stress tensor. This definition applies well to granular materials, as the stresses that act on the boundaries of the systems are transmitted through the constituent components, and assuming the system is in quasi-static equilibrium, the boundary stresses are equal to the internal stresses.

We then calculate the bulk friction coefficient $\mu$ using the stress tensor 
\begin{equation}
    \mu = \frac{\tau}{P} = \frac{ 2 \sqrt{(\sigma_{11} - \sigma_{22})^2  + 4 \, \sigma_{12}\sigma_{21}}}{(\sigma_{11} + \sigma_{22})}.
    \label{eq:mu}
\end{equation}
The bulk friction is the ratio of shear $\tau$ to 
 the mean stress $P$ and can be thought of as the resistance to deformation of the packing as a whole. We note that even assemblies of frictionless particles have non-zero bulk friction coefficients because of volume exclusion effects~\cite{hohler_many-body_2017, roux_geometric_origin_2000, peyneauFrictionlessBeadPacks2008, azemaInternalFrictionAbsence2015}. 
The mean stress $P$ is given by the trace of the granular stress tensor: $P = (\sigma_{11} + \sigma_{22})/2$. The shear stress is the summation of the difference between principal stress directions and the off axis stress components $\tau = \sqrt{(\sigma_{11} - \sigma_{22})^2  + 4 \, \sigma_{12}\sigma_{21}} $. 

The stress tensor and the bulk friction coefficient can be measured at either the microstructural scale or the boundary scale. Although accurate, the micromechanical model requires knowledge of every particle and contact. One way to think about the Stress-Force-Fabric relation is as a method to reduce the amount of information required to understand the bulk response.

\subsection{Stress-Force-Fabric relation}
\label{SecSFF}

The Stress-Force-Fabric (SFF) relation reconsiders the micromechanical stress tensor (Eq.~\ref{eq:stresstensor}) by grouping the contact and force vectors based on the contact normal vector angle $\theta$. The stress tensor is expanded using Fourier expansions to map the magnitude of directionality of each term, and relates the bulk tensor to anisotropy in forces and fabric. Here we summarize the derivation described in previous work~\cite{rothenburg_analytical_1989,li_fabric_2014,li_stressforcefabric_2013}, and rewrite Eq.~(\ref{eq:stresstensor}) such that the summation is grouped by contact normal vector that can be described by a generic vector $\vec{n} = (\cos \theta ;\sin \theta)$ that corresponds to angle $\theta$
\begin{equation}
    \sigma_{ij} =\frac{1}{S} \sum_\theta \left< \ell_i f_j \right> ({\theta}) \Delta M (\theta)
    \label{eq:anglesum}
\end{equation}
where $\vec \ell$ is the branch vector which joins the center of two contacting particles, $\left< \right>(\theta)$ indicates the average of the value within the brackets as evaluated for all contacts that are within an angular element $\Delta\theta$ centered at angle $\theta$. We note a subtlety in notation, where in Eq.~\ref{eq:stresstensor} the summation runs over all particles while it runs over all contacts in Eq.~\ref{eq:anglesum}. The average outer product between the branch vector and the force vector is multiplied by the number of contacts  $\Delta M(\theta)$ that fall within the angular element when it is centered about $\theta$.

Equation~\ref{eq:anglesum} can be generalized by considering the contact normal probability distribution function (pdf)
 $E^c(\theta) = \Delta M(\theta) / \Delta \theta$.  
When taken in the thermodynamic limit, the summation turns into an integral over infinitesimally small angle, d$\theta$ and the stress tensor may be written as
\begin{equation}
\sigma_{ij} = \frac{N_c}{S} \oint_\theta E^c(\theta) \left< \ell_i f_j \right>(\theta) d\theta,
\label{eq:stressexpand}
\end{equation}
with $N_c$ the total number of contacts in the assembly. From here, we outline three simplifications.

\paragraph*{Simplification 1} The branch vectors $\vec \ell$ and force vectors $\vec f$ are statistically independent, and thus
\begin{equation}
\left< \ell_i f_j \right>(\theta) = \zeta \left< \ell_i\right>(\theta) \left< f_j\right>(\theta) 
\label{Eq_ksi}
\end{equation}
where $\zeta$ is a direction-independent scalar that is typically of order 1~\cite{li_fabric_2014}. We tested this assumption with our experimental data and found that there was no directional dependence between the two vectors and can use the right hand side of the equation with $\zeta = 1$ (See SI).
Moreover, for circular grain shapes with small size variation, $\vec{\ell} = d \hat{n}$, with $d$ the mean size diameter, and therefore Eq.~\ref{Eq_ksi} becomes:
\begin{equation}
\left< \ell_i f_j \right>(\theta) = d \; \left< f_j\right>(\theta) \; n_i(\theta) 
\label{Eq_ksi_2}
\end{equation}

\paragraph*{Simplification 2} The contact normal pdf $E^c(\theta)$ can be expanded into a tensorial Fourier expansion up to the second term  
\begin{equation}
    E^c(\theta) = \frac{1}{2\pi} \left[ 1+ a 
    \begin{pmatrix}
    \cos{\theta_c} & \sin{\theta_c}\\ \sin{\theta_c} & -\cos{\theta_c}
    \end{pmatrix}\begin{pmatrix}
    \cos{\theta} \\ \sin{\theta} 
    \end{pmatrix}\right],
\end{equation}
where $a$ corresponds to the Fourier coefficient or the magnitude of directional dependence and $\theta_c$ corresponds to the principal direction of the largest contact density. The contact normal pdf can therefore be approximated as
\begin{equation}
     E^c(\theta) = \frac{1}{2\pi} \left[ 1+ a \cos 2(\theta - \theta_c )\right].
    \label{eq:contacts}
    \end{equation}
This expansion results in an isotropic term summed with an anisotropic term. 
As it will be discussed in more detail in Sec.~\ref{Sec_Results}, we show that the approximation given by Eq.~\ref{eq:contacts} is valid in all four loading experiments.

\paragraph*{Simplification 3} 
The contact force vector $\vec f$ can be split into its normal and tangential components, and thus the mean angular distribution of the total force is rewritten in terms of the mean angular distribution of normal forces $\langle f_n\rangle (\theta)$ and the  mean angular distribution of tangential forces
$\langle f_t\rangle (\theta)$: $\langle f_j\rangle(\theta) = \langle f_n\rangle (\theta)n_j(\theta) + \langle f_t\rangle (\theta) t_j$, where $\vec{t}$ is an orthonormal unit vector oriented along the tangential force. As for $E^c(\theta)$, these two distributions can be then approximated using a first order Fourier expansion as
\begin{equation}
    \langle f_n\rangle(\theta) = f_0\left(1+a_n \cos2\left(\theta - \theta_f\right)\right),
    \label{eq:normal}
\end{equation}
\begin{equation}
    \langle f_t\rangle(\theta) = f_0 \; a_t \sin2(\theta - \theta_f),
    \label{eq:tangential}
\end{equation}
where $f_0$ is the mean normal force, $a_t$ and $a_n$ are the anisotropy amplitudes for the tangential and normal forces, respectively. $\theta_f$ is the angular direction of the largest average force. The anisotropy in normal forces $a_n$ is related to the force transmission through the granular material, where the anisotropy of the tangential forces relates heterogeneous directions of friction mobilization~\cite{azemaForceChainsContact2012,azemaPackingsIrregularPolyhedral2013}.

\paragraph*{The Stress Rule and Sum Rule}
With these three simplifications, requiring symmetry in the Cauchy stress tensor ($\sigma_{12}= \sigma_{21}$) for a two dimensional system, and assuming the principal direction is the same for forces and contacts, $\theta_f = \theta_c = \theta_a$, the stress tensor can be written as a {\it Stress Rule}:
\begin{multline}
\sigma_{ij} = \frac{z \nu}{\pi d } f_0\, \Biggl[\left(1+\frac{aa_n}{2}\right) \delta_{ij}\\
+ \frac{1}{2} \left(a+a_n+a_t\right)\begin{pmatrix}\cos{2}\theta_a & \sin{2}\theta_a\\\sin {2}\theta_a & -\cos{2}\theta_a\end{pmatrix}\Biggr],
\label{eq:stressrule}
\end{multline}
where we substitute the area $S = N \pi d^2 / 4\nu$, with $\nu$ being the packing fraction. This expansion of the stress tensor may be used to calculate the bulk friction coefficient $\mu$. Inserting Eq.~\ref{eq:stressrule} into Eq.~\ref{eq:mu}, and neglecting cross product between anisotropies, results in a simple {\it Sum Rule}:
\begin{equation}
    \mu = \frac{1}{2}(a+a_n+a_t),
    \label{eq:SFF}
\end{equation}
that was first reported in \citet{rothenburg_analytical_1989}. It directly relates the bulk friction coefficient to the anisotropy amplitudes, indicating that the resistance to deformation in a granular material simply depends on the amount of anisotropy that is present in the granular assembly. 
In the analysis that follows, we will test the validity of the Stress Rule (Eq.~\ref{eq:stressrule}) and Sum Rule (Eq.~\ref{eq:SFF}) for each of the four loading geometries.

\section{Results}
\label{Sec_Results}

 \subsection{Measuring angular distributions} 
\begin{figure*}
    \centering
    \includegraphics[width=1\textwidth]{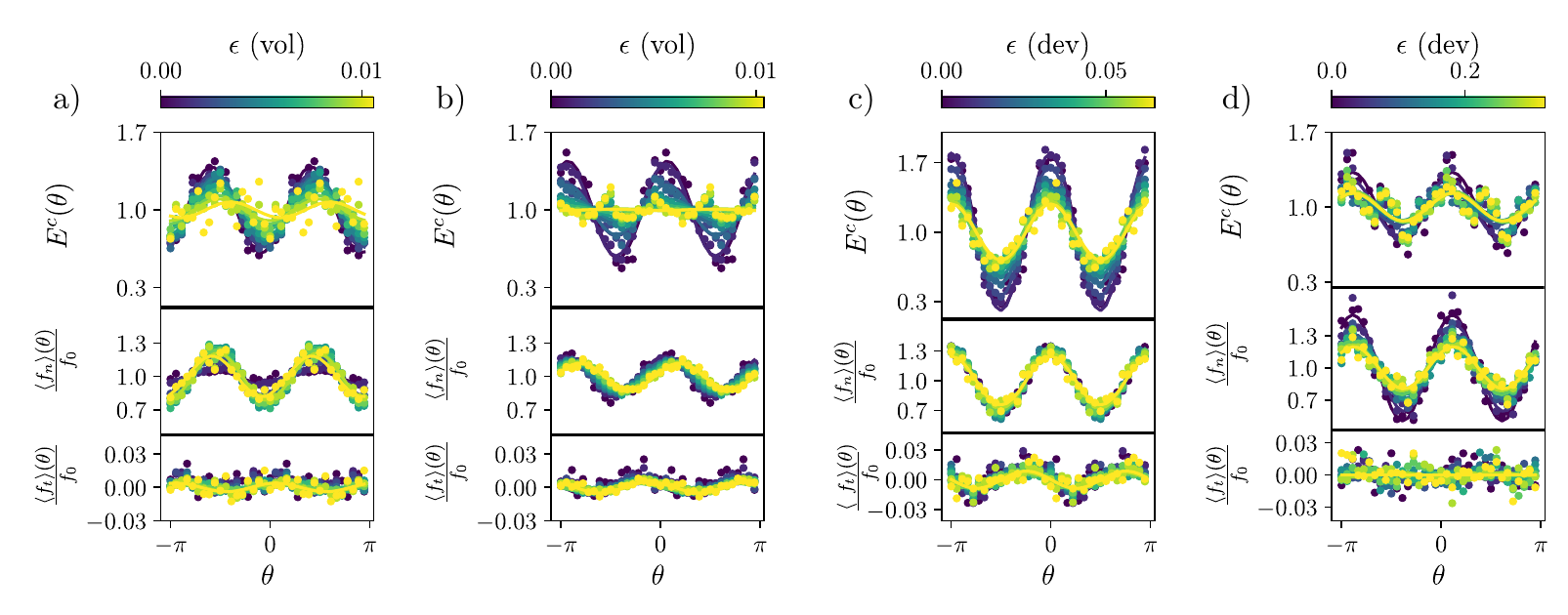}
   \caption{Top row: angular distributions of the contact normals $E^c(\theta)$; Middle row: mean normalized normal forces $\langle f_n\rangle(\theta)$; Bottom row:  mean normalized tangential forces $\langle f_n\rangle (\theta)$. Each distribution is plotted for the four loading histories (a) uniaxial, (b) isotropic, (c) pure shear and (d) annular shear. For the three rectangular geometries (a-c), $\theta$ is measured relative to the system-wide $+x$ axis, while for the annular geometry it is measured relative to the radial axis $R$. The increments of volumetric or deviatoric strain for each loading history are (a) $\epsilon = [0, 0.013]$, (b) $\epsilon = [0, 0.013]$ (c) $\epsilon = [0, 0.06]$ and (d) $\epsilon = [0, 0.3]$ in 10 evenly spaced intervals, with the value indicated by color ranging from low (dark blue) to high (light yellow).}
    \label{fig:angular_stribution}
\end{figure*}

Figure \ref{fig:angular_stribution} shows the evolution of the different angular distributions $E^c(\theta)$ (top row), $\langle f_n \rangle(\theta)$ (middle row) and $\langle f_t \rangle(\theta)$ (bottom row) for each of the loading geometries as the packing is under strain (shown via the colorbar).
For videos of the co-evolution of the positions, force networks, and angular histograms, see the SI videos available for each of four loading geometries. For the three rectangular geometries, $\theta$ is taken relative to the positive $x$ coordinate, while for the annular geometry it is taken relative to the radial coordinate $R$.

In general, all distributions have the same shape, with a pronounced peak around a preferred direction, depending on the loading geometries tested. In the initial state, the shape of the distributions is the result of the sample construction and loading protocol. During deformation, the global shape of the distributions is preserved and only the amplitude around the preferred direction varies.  In agreement with the hypothesis formulated earlier, the preferred directions of these three distributions, for a given geometry, are globally identical (i.e. $\theta_c=\theta_f$).

More specifically, in the case of uniaxial compression (a) and isotropic compression (b), the peaks observed on $E^c(\theta)$ attenuate until the distribution becomes almost isotropic $E^c(\theta)\sim 1$ in both cases (i.e. no preferred contact orientation is observed in the densest state). The initially isotropic shape of the normal force distribution $\langle f_n\rangle(\theta)$ becomes anisotropic and oriented along the loading direction in case (a) and, as expected, remains globally isotropic in case (b). In cases (a) and (b) the systems are compressed, therefore, the deformation stops when the contact network and the force network reach equilibrium.

In the pure shear configuration, the contacts and forces are mainly oriented along the $\hat{x}$ direction (as clearly observed in Fig.~\ref{fig:loading}(c) and therefore the peaks of the distributions are well along the $\theta \sim 0$ direction.  Finally in the case of annular shear the principal orientation of contacts and forces is approximately along the $\theta\sim\pi/4$, analogous to the case of planar shear~\cite{da_cruz_rheophysics_2005}. The peaks on the contact and normal force distributions decrease slightly during deformation, until the distributions becomes, on average, constant at large strain.

Finally, in each geometry, the distribution of tangential forces shows  small variations around $0$, consistently with previous results obtained with Discrete Element Simulations~\cite{binaree_combined_2020}.
The distributions also show peaks that are slightly offset from those of the contact and normal force distributions, indicating the preferred directions in which friction is predominantly mobilised~\cite{azemaForceChainsContact2012, azemaPackingsIrregularPolyhedral2013}.

Regardless of all the subtle variations in the shape of the distributions we have described, the main result is that our experiments show a regular behaviour with $\pi$ periodicity of the angular distributions of the contacts and forces distributions. Thus, as shown in Fig~\ref{fig:angular_stribution}, our experimental data for each distribution are well described by Eq.~\ref{eq:contacts}, Eq.~\ref{eq:normal} and Eq.~\ref{eq:tangential} at increasing levels of strain.

\subsection{Texture and anisotropies}

The various changes in the shape of the angular distributions discussed in the previous section can be described more quantitatively by the evolution of the anisotropic amplitudes $a$, $a_n$ and $a_t$ used to fit Eq.~\ref{eq:contacts}, Eq.~\ref{eq:normal} and Eq.~\ref{eq:tangential} respectively.  In addition to these parameters, Eq.~\ref{eq:stressrule} also reveals that any analysis of the fabric and force network must be complemented by an analysis of the connectivity of the contact network (i.e. the coordination number $z$), the packing density (i.e. the solid fraction $\nu$) and the mean normal force  $f_0$.

\begin{figure*}
    \centering
    \includegraphics[width=1\textwidth]{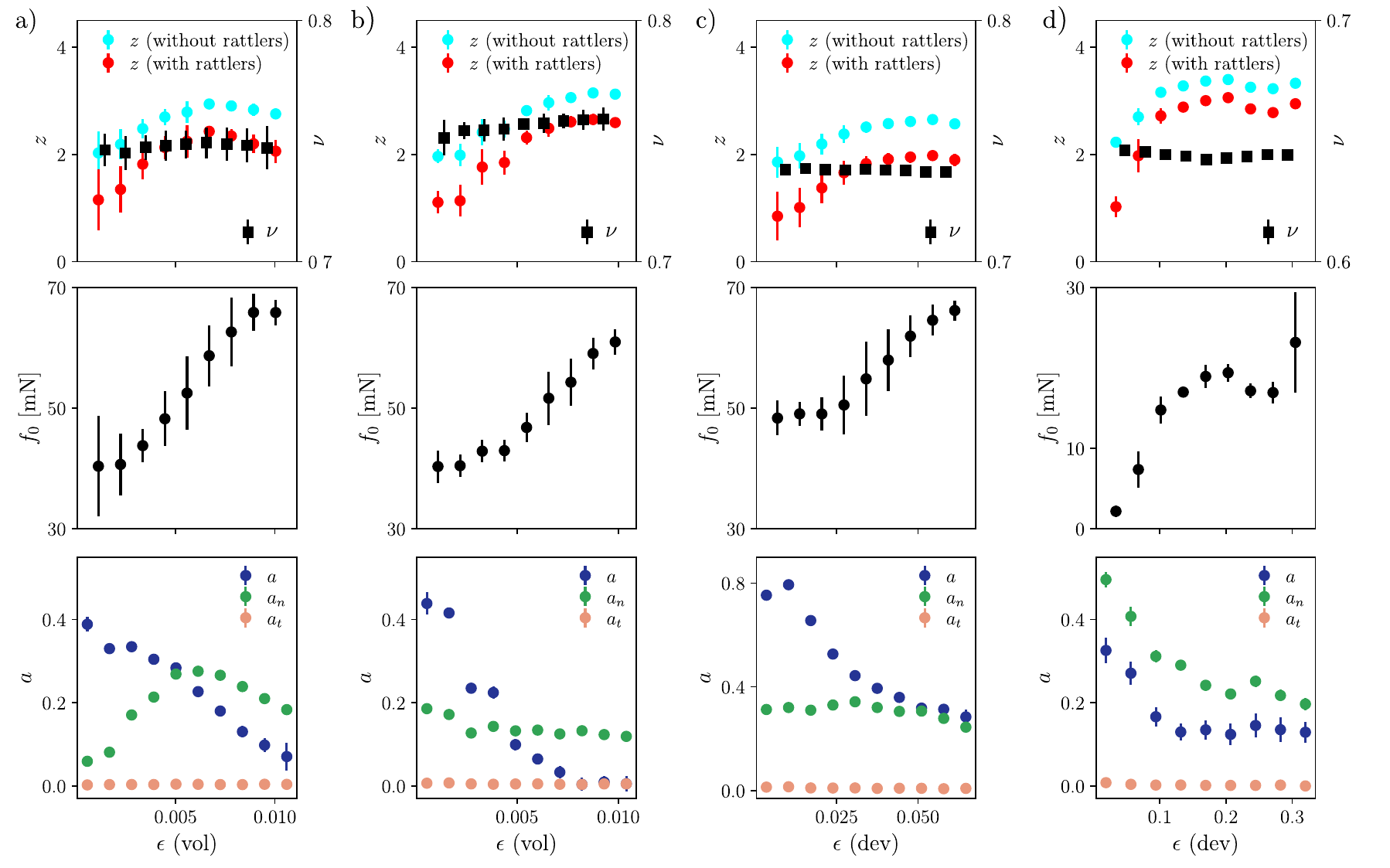}
   \caption{We show the development of particle-scale parameters as a function of strain for columns (a) uniaxial compression (b) isotropic compression (c) pure shear (d) annular shear. Top row: The variation of the coordination number $z$, calculated with and without rattlers and the packing fraction $\nu$ as a function of strain $\epsilon$. Middle Row: The average normal contact force in the material as a function of strain $\epsilon$. Bottom row: The anisotropy magnitudes $a$, $a_n$ and $a_t$ at differing levels of strain $\epsilon$.}
    \label{fig:microstructure}
\end{figure*}    

The top row of Fig.~\ref{fig:microstructure} shows the evolution of both coordination number $z$ and packing density $\nu$ for all four loading geometries (a)--(d),  averaged over all trials, as a function of strain $\epsilon$. We distinguish the evolution of $z$ computed without (blue) or with (red) rattlers (i.e. particles with less than 2 contacts) and use $z$ without rattlers in
Eq.~\ref{eq:stressrule}. Initially, the values of $z$ are quite low ($z\approx 2$ measured without rattlers and $z \approx 1$ with rattlers) for all geometries tested. We observe $\nu \approx 0.73$ to $0.75$ in the rectangular geometries (a)--(c), consistent with our prior results in this apparatus~\cite{liu_spongelike_2021}, but as low as $\nu = 0.64$ in the sheared annulus (d). This low value arises from significant exclusion effects near the two rough walls and heterogeneity in the packing while under shear. Under loading, $\nu$ remains globally constant (trivially in column (c), as the pure shear test is performed to conserve area). This behavior was as also reported in numerical discrete element studies of sheared frictionless (or weakly frictional, as in our experiments) grain packing~\cite{peyneauFrictionlessBeadPacks2008, azemaInternalFrictionAbsence2015}. 

In concert, $z$ increases under loading for all geometries, tending towards a constant value close to $3$ (for $z$ without rattlers) at the highest deformation. This plateau is expected because in cases (a) and (b) the systems tend towards a static mechanical equilibrium, which is quickly reached typically for $\epsilon\sim 0.01$. In cases (c) and (d), the systems are sheared at constant pressure with (d) sheared at a constant velocity and therefore also tend towards a steady-state equilibrium, also known as the critical state in soil mechanics. The critical state is equivalent to the dilatancy transition~\cite{Schroter_phase_2007}. This plateau is reached at $\epsilon\sim 0.04$ for pure shear (c) and at $\epsilon\sim 0.15$ for annular shear (d). Finally, it should be noted that the value of $z$ (without rattlers) obtained at static or quasi-static equilibrium is less than $3$, the value expected for frictional systems such as ours; similarly low values have be observed in simulations~\cite{vanderNaald_minimally_2024} and well as previously in this apparatus~\cite{liu_spongelike_2021}. One contributing factor is the reduction in $z_c$ from the boundary/finite-size effects~\cite{perrin_nonlocal_2021, wyart_geometric_2005, ellenbroek_jammed_2009}, but additionally our thresholding to find contacts systematically under-counts (see SI). 

The middle row of Fig.~\ref{fig:microstructure} reports the evolution of the mean normal contact force $f_0$ for all four loading geometries (a)--(d), averaged over all trials,  as a function of strain $\epsilon$. We observe that  $f_0$ increases with $\epsilon$, even as  $z$ remains  constant in the rectangular loading geometries (a)--(c). This means that the contacts support increasingly strong forces under the effects of increased strain. In contrast, for the angular shear geometry (d), the value of $f_0$ saturates in accordance with  $z$, corresponding to the fact that the flow reaches a steady state.

Finally, the bottom row of Fig.~\ref{fig:microstructure} shows the evolution of the three anisotropic parameters ($a$, $a_n$ and $a_t$) for all four loading geometries, averaged over all trials, as a function of the strain $\epsilon$. First, we note that in the initial state ($\epsilon=0$), $a$ is near $0.4$ in the three of the geometries (a),(b), and (d) as high as to $0.75$ in (c). This means that our particle-placement protocols already induce a preferential orientation of the contacts. Upon loading,  in all cases, $a$ first decreases with $\epsilon$. When considered in concert with the observed increase in $z$, this highlights that more contacts are created perpendicular to the main shear direction than parallel to it. Once $z$ becomes constant, later in the loading process, different behaviours are obtained depending on the geometry tested. For geometries (a) and (b), $a$ continues to decrease, and reaches a value close to $0$ for both in the large-strain limit. This reflects the fact that in this state the particles tend to reorganise without forming new contacts. In contrast, for geometries (c) and (d), $a$ tends to a plateau with an approximate value of $0.25$ for case (c) and $0.19$ for case (d), implying that the initial anisotropic nature of the contact network is preserved, albeit less pronounced.

In general, the evolution of the normal force anisotropy $a_n$ does not necessarily follow the same evolution as the contact anisotropy $a$. In case (a), $a_n$ first increases with $\epsilon$ consistently with the loading direction leading to the birth of vertical force chains (and the increases of $f_0$. Then $a_n$ passes through a peak coinciding with the saturation of $z$, before decreasing slightly towards a value of $0.2$ associated with the reorganisation of the particles in this last phase. In cases (b) and (c), $a_n$ remains globally constant with $\epsilon$, while $a$ decreases. These two cases have to be analysed separately. In (b), the fact that $a_n$ does not tend to $0$, while the compression is isotropic is almost certainly due to the small size of the system, which partially retains the memory of the initial state. In (c) the system is sheared and the force chains adapt to the applied stress until the residual state is reached. A similar mechanism is at work in case (d), where $a_n$ decreases and then tends towards a plateau in the residual state. Finally, $a_t$ is very close to $0$ in all of the tests, because the friction between the grains is low.

\subsection{Evaluating the bulk response via the SFF}

\begin{figure*}
    \centering
    \includegraphics[width=1\textwidth]{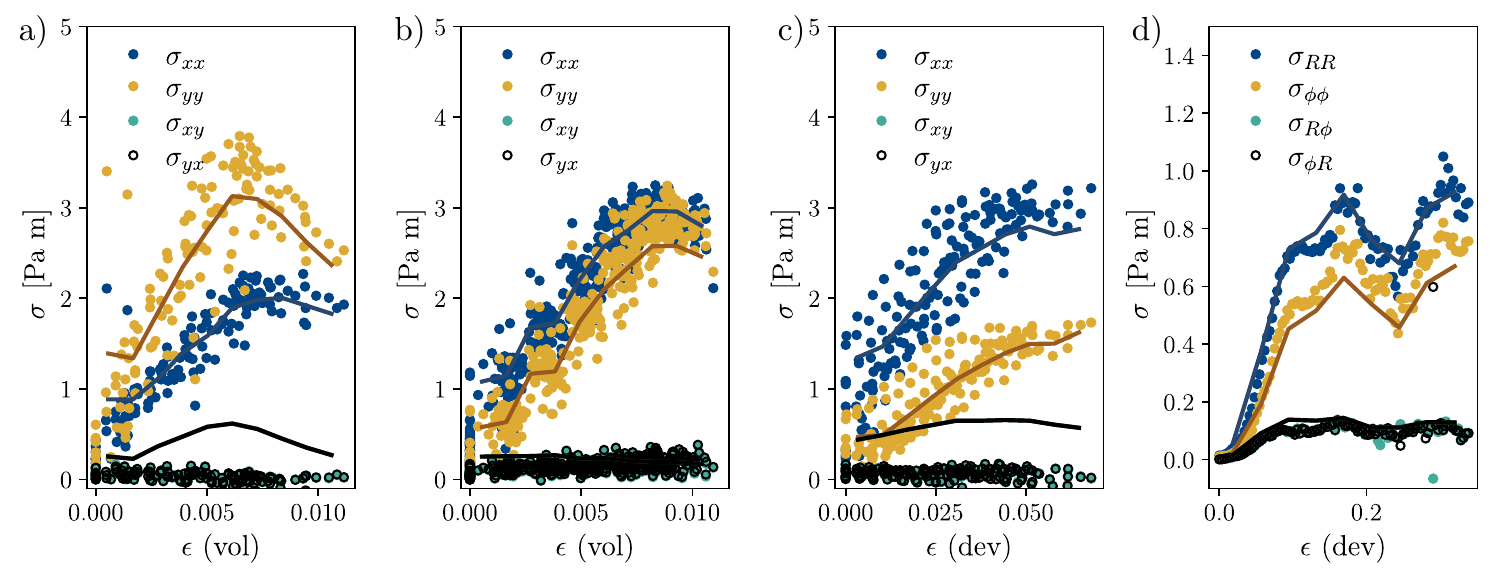}
   \caption{ Evolution of the stress tensor components $\sigma$ at different levels of strain $\epsilon$ for the four loading geometries, uniaxial compression  (a), isotropic compression (b), pure shear (c), and  annular shear (d). The rectangular geometries follow Cartesian principle stress directions, where the annular shear cell follows polar coordinates. The predicted values based on the stress rule (Eq.~\ref{eq:stressrule}) are shown as solid lines.}
    \label{fig:stressrule}
\end{figure*}

Next, we consider the macroscopic behaviour of each loading geometry, through the evolution of the components of the stress tensor under strain. We test the Stress Rule (Eq.~\ref{eq:stressrule}) and thereby propose a more detailed view of the role of the fabric and force network on bulk properties.
Fig.~\ref{fig:stressrule} shows the four stress components calculated according to Eq.~\ref{eq:stresstensor} and averaged over all trials for each geometry (a)--(d). 

In all cases, the diagonal terms of the stress tensor ($\sigma_{\alpha \alpha}$, where $\alpha$ is $x$, $y$, $R$ or $\phi$) increase first with strain, and then tend to plateau in the dense equilibrium state (a) and (b) or in the residual quasi-static state (c) and (d). For each loading, the relative magnitude of the two diagonal terms is consistent with the given loading geometry. For uniaxial loading (a), $\sigma_{xx}< \sigma_{yy}$, which is consistent with the loading being applied along the $\hat{y}$ direction, while for pure shear (c), $\sigma_{xx}> \sigma_{yy}$, indicating stress along the $\hat{x}$ direction. In isotropic compression (b), $\sigma_{xx}$ is naturally very close to $\sigma_{yy}$ due to the packing being isotropically loaded. Finally, for annular shear (d), $\sigma_{rr}> \sigma_{\theta\theta}$ due to the continuous imposed shear. In contrast, the deviatoric terms of the stress tensor remain relatively constant throughout the deformation and are much smaller than the diagonal terms. which are equal to each other, yet confirming the symmetry of the stress tensor

The Stress Rule approximations (Eq.~\ref{eq:stressrule}) are shown in solid lines in Fig.~\ref{fig:stressrule}. We observe that its predictions hold for nearly all stress components, and throughout the evolving deformation. The weakest agreement arises in the deviatoric terms ($\sigma_{xy}, \sigma_{yx}$) for pure shear (c) and uniaxial compression (a), where the corresponding Eq.~\ref{eq:stressrule} tend to overestimate the measured results. Based on these observations, Eq.~\ref{eq:stressrule} provides an explanation for the evolution of macroscopic stresses under strain.

\begin{figure*}
    \centering
    \includegraphics[width=1\textwidth]{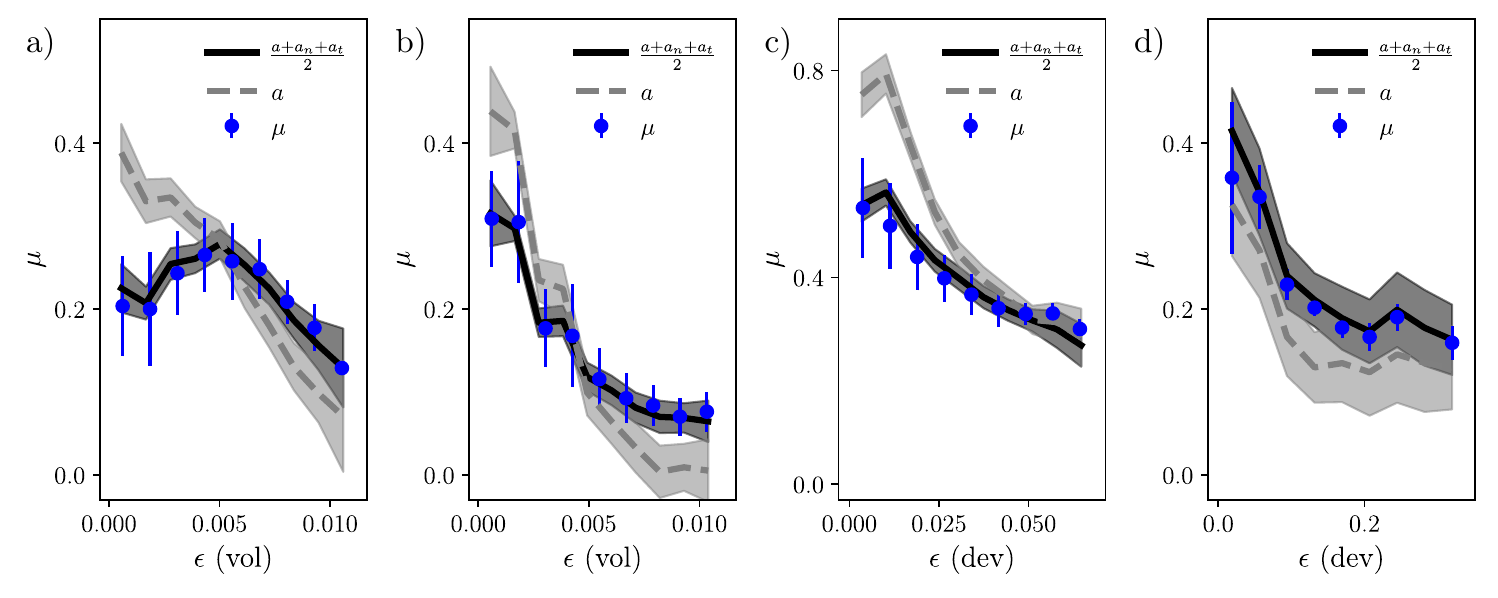}
   \caption{Evolution of the  bulk friction coefficient $\mu$ as a function of strain $\epsilon$ for the four loading geometries, uniaxial compression (a), isotropic compression (b), pure shear (c), and annular shear (d). We compare the measured bulk friction coefficient from eq.~\ref{eq:mu} (dots) to the prediction from the stress-force-fabric relation sum rule from eq.~\ref{eq:SFF} (black line). In addition, we show the prediction of using only contact information simplifying eq.~\ref{eq:SFF} to $\mu \sim a$ (grey dashed line). }
    \label{fig:sumrule}
\end{figure*}

The sum of the anisotropies defines the bulk friction coefficient $\mu$ (see Eq.~\ref{eq:SFF}). Thus, Fig.~\ref{fig:sumrule} shows the variation of $\mu$, averaged over all tests, in each of the geometries (a--d), measured directly from the components of the stress tensor (filled blue circles) and predicted by the anisotropies Eq.~\ref{eq:SFF}. As we can see, the prediction is perfect throughout the whole deformation, thus providing the very first experimental validation of the SFF across various shear geometries. Under uniaxial shear (a), $\mu$ first increases slightly from $0.2$ to $0.3$ (the gain in $a_n$ offsets the loss in $a$) before decreasing to a value close to $0.1$. This final value results from decreases in both $a$ and $a_n$. Under isotropic compression (b), there is no shear and $\mu$ decreases from $0.4$ to $\sim 0.1$, emphasised by the decrease of both $a$ and $a_n$. The fact that $\mu$ does not reach $0$ in the steady state results from the persistence of anisotropy in the normal forces, probably induced by the effect of the size of the systems. In (c) and (d) $\mu$ decreases to a plateau value at about $\mu\sim0.2$, mainly due to the decrease of $a$ in (c) and of both $a$ and $a_n$ in (d). This result confirms that anisotropic structures in force and fabric control bulk friction and resistance to different types of strain in granular materials. 

These findings lead us to a key application of the SFF framework: to improve the characterisation of particulate systems for which interparticle force measurements are inaccessible. Following the observations (see Fig.~\ref{fig:microstructure}(a)-(d) $a(\epsilon)$ plots) that (1) $a \approx a_n$ in many scenarios and (2) $a_t \ll a_n$, we are motivated to consider the utility of the sum rule using only contact information (i.e., the contact anisotropy $a$). 
In Fig.~\ref{fig:sumrule}, the dashed lines are calculated using the approximation $a_n = a$ and $a_t = 0$, and are still semi-quantitatively predictive of $\mu$ trends. Thus, in the absence of force data, anisotropy in contact distributions are a reasonable proxy for determining bulk properties, including the onset of strain softening, at least in circular and nearly mono-size particle assemblies. We note this implies that  at large strains, the force and fabric transmit load coaxially and with similar contributions.  For small strains, this approximation breaks down: we instead see large errors (up to 50\% of $\mu$) which suggests that at low strains there is a significant difference between the contact anisotropy and the force anisotropy. We note that this approximation may be broken for highly polydisperse assemblies~\cite{azema_shear_2017,cantorMicrostructuralAnalysisSheared2020} or assemblies of any shape~\cite{azemaNonlinearEffectsParticle2012,azemaStressstrainBehaviorGeometrical2010,azemaPackingsIrregularPolyhedral2013}, and would require a separate investigation to pursue in more detail.

To gain a physical intuition into the dominant origins of the stress tensor, we return to Eq.~\ref{eq:stressrule}, where we can simplify the expansion to first order by neglecting the variations in the cross product of the anisotropies (i.e., the term $a a_n/2$ in eq.~\ref{eq:stressrule}), and consider that the packing fraction $\nu$ remains approximately constant during the deformation. Therefore, the evolution of the normalized components of the stress tensor with $\epsilon$ is governed by the evolution of $z$, the average normal force $f_0$, and the evolution of the bulk friction coefficient $\mu$, through the Sum Rule $(a+a_n+a_t)/2$. We deduce that the increase in $\sigma_{\alpha\alpha}$ with $\epsilon$ (where $\alpha$  stands alternatively for $\{x,y,R,\phi\}$) is essentially due to the increase in $z$ and $f_0$, which dominates the bulk response of the stress tensor. However, after equilibrium is reached and the stress tensor plateaus, we see that the increase due to $f_0$ is balanced by a decrease in the bulk friction coefficient: this mechanism may be a fresh way to understand shear-weakening.

\section{Conclusions}

In this study, we have shown experimental validation of the SFF relation using photoelastic particles in four different loading geometries. We track the development of the coordination number, average force, and packing fraction as a function of strain along with anisotropy in the distribution of fabric and forces. The success of the SFF relation indicates that the bulk friction coefficient of a granular material arises due to a combination of anisotropy in both fabric and forces. As well, we compare the role of the anisotropies with other common characteristics and show that considering the contribution of the anistropies is necessary to fully capture the evolution of the stress tensor with increasing strain. Although not considered here, we presume the shear modulus should follow a similar dependence on the evolution of the coordination number~\cite{zaccone_shear_2011}. Importantly, the SFF relation remains valid even under different geometries and loading geometry beyond the system for which the framework was initially developed. We show that there are similarities how the anisotropy amplitudes of the two compressive loadings develop under strain, something to be expected, as the uniaxial case is an extension of the isotropic compression. In addition, we have shown that the first order Fourier expansion of the  stress tensor is sufficient to described bulk properties, like the bulk friction coefficient and the stress tensor components. 

We note that this framework does not address the role of mesoscale structure and therefore does not encompass prediction of rigidity~\cite{liu_spongelike_2021}; however, the SFF relation is a promising pathway to address the open challenge of connecting fluid-like granular theories~\cite{kamrin_nonlocal_2012, jop_constitutive_2006, berzi_granular_2024} to quasi-static granular behaviour, and has been shown to capture even transient dynamics~\cite{lee_preprint}.  Finally, the framework may additionally provide a way to link fabric anisotropy to bulk friction in experimental systems for which vector contact forces cannot be measured, as illustrated by the success of the contact anisotropy $a$ alone at large strains. We note that in the absence of force information, the geometric anisotropy provides a good estimate of bulk friction, at least when circular particles are considered, and could be easily applied to other deformable assemblies, including cellular tissues, foams, dense colloid suspensions and rocks with grain boundaries.

\begin{acknowledgments}
{\bf Acknowledgments:} The authors thank Jonathan Bar\'es and Tejas Murthy for fruitful discussions, and Ben McMillan for improvements to the PEGS photoelastic solver. The authors thank the Lorentz Center for hosting the workshop Getting into Shape where the idea for this paper was formed. C.L.L. acknowledges funding from the NSERC postdoctoral fellowship, and K.E.D and C.L.L from the National Science Foundation (DMR-2104986).
\end{acknowledgments}

\end{document}


\title{Loading-dependent microscale measures control bulk properties in granular material: an experimental test of the Stress-Force-Fabric relation}
\author{Carmen L. Lee}
\affiliation{Department of Physics, North Carolina State University, Raleigh, North Carolina, 27695, USA}

\author{Ephraim Bililign}%
\affiliation{Department of Physics, North Carolina State University, Raleigh, North Carolina, 27695, USA}
\affiliation{James Franck Institute and Department of Physics, University of Chicago, Chicago, IL, USA}

\author{Emilien Az\'ema}
\affiliation{LMGC, Universit\'e de Montpellier, CNRS, Montpellier, France}
\affiliation{Department of Civil, Geological, and Mining Engineering, Polytechnique Montréal, Montréal, Canada.}
\affiliation{Institut Universitaire de France (IUF), Paris, France}

\author{Karen E. Daniels}
\affiliation{Department of Physics, North Carolina State University, Raleigh, North Carolina, 27695, USA}

\maketitle

\section{Contact detection}
\begin{figure}[b]
    \centering
    \includegraphics[width=\columnwidth]{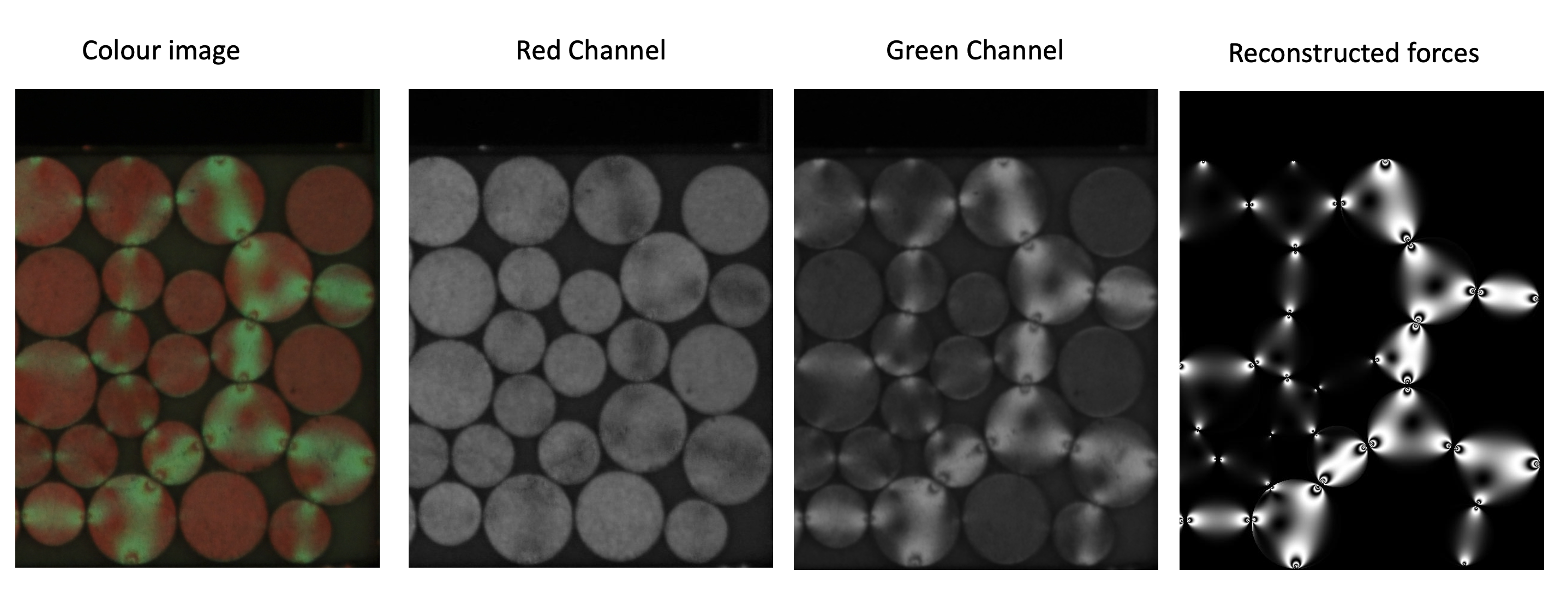}
    \caption{A series of images showing the image analysis pipeline in the PhotoElastic Grain Solver (PeGS). (a) The original colour image shows particles lit with red, unpolarized light, and green polarized light. (b) The red channel of the original image is used to track the position of the particles using a Hough Circle transform to detect the particle position and radius. (c) The green channel of the original image shows the forces on each particle. The algorithm detects the contacts between particles and between the particles and the wall. The algorithm then solves for the forces due to each contact. (d) A reconstructed image of the particles is created to verify the accuracy of the solved forces. }
    \label{fig:photoelasticity}
\end{figure}

To find the contact vectors and vector forces, we follow the image processing pipeline laid out in the Supplementary Material in \citet{liu_spongelike_2021}. The particles are lit with red, unpolarized light to image normally and green circularly polarized light to form a polariscope. The particle are detected using the red channel in the camera, and the forces are detected using the green channel (see Fig.~\ref{fig:photoelasticity}).

In darkfield photoelastic imaging, traditional method calculate the local stress at the potential contact locations using intensity gradients. At a potential contact location, the local intensity gradient squared ($G^2$) is calculated in a circular region of interest. In the PhotoElastic Grain Solver code ~\cite{kollmer_photo-elastic_nodate}, above a threshold value the contact is accepted as a force bearing contact, otherwise it is discarded as non-force bearing. While this method is highly robust in identifying regions with large photo-responses, it has a low noise tolerance and artificially high thresholds must be set to reduce the number of false positive contacts. As a result, small contacts are missed and systematically excluded from analysis.

\begin{figure}
    \centering
    \includegraphics[width=0.65\columnwidth]{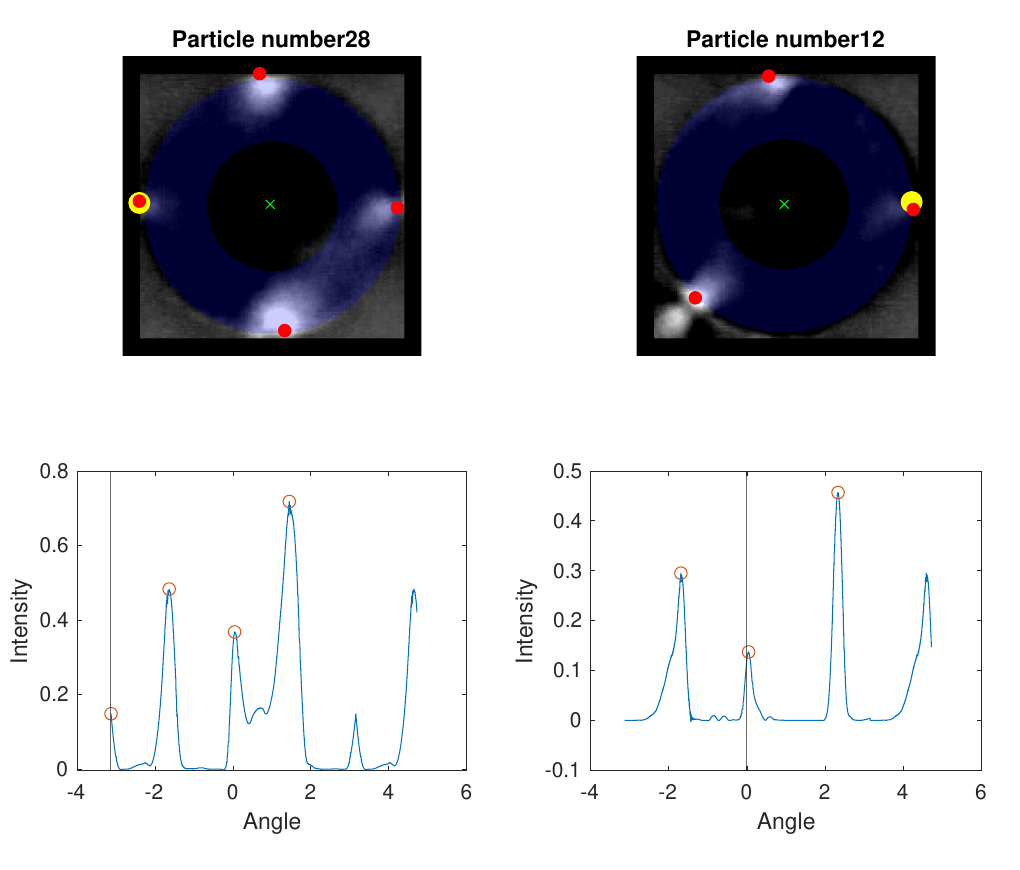}
    \caption{A sample of two neighbouring particles that pass the peak finding algorithm. The top panel shows the particles overlaid with the annular area over which the intensity profile is averaged (blue, color online). The red dots indicate the peaks found by the peak finding algorithm, and yellow larger dot indicates the nominal contact position of the other particle. The lower panels show the intensity profiles as a function of angle from the positive horizontal direction. The black vertical line indicates the nominal contact angle.}
    \label{fig:peakfinding}
\end{figure}

In our analysis for this paper, we implement two main changes. The first is a change to the force solving algorithm proposed by McMillan to fix an error in solving the tangential forces~\cite{McMillan_note}. The second change is we the implementation of a secondary step to access contacts with weak forces that would ordinarily fall below the $G^2$ threshold. Boosting the contrast of the image, we are able to resolve faint contacts with $G^2$ values below the contact detection threshold but above the noise floor. These are visible when taking an angular intensity profile around the outer ring of the particle. Figure~\ref{fig:peakfinding} shows an example of two particles that are in weak contact with each other. The top panels show the two particles, with the blue ring indicating the area over which the intensity profile is taken.  We then implement the peakfinding algorithm in \textsc{MATLAB} to find intensity peaks. If there is a peak that is aligned with the expected location of the contact for both particles, within tolerance, it is then considered a contact. 

\begin{figure}
    \centering
    \includegraphics[width=0.65\columnwidth]{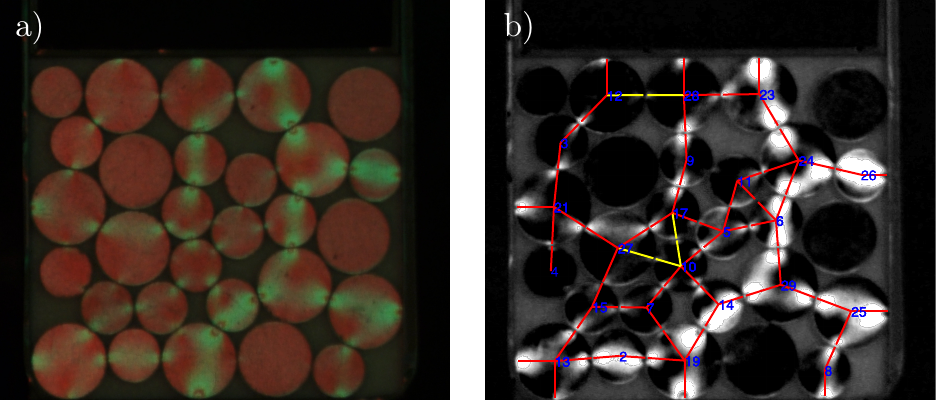}
    \caption{a) A sample image of photoelastic particles. b) The image is shown with increased contrast green channel to illustrate the small forces. Overlaid on the image is the detected contact network. Red lines indicate contacts found using the $G^2$ threshold and yellow lines show the supplementary contacts found using the peak finding method.}
    \label{fig:contactnetwork}
\end{figure}

\begin{figure}
    \centering
    \includegraphics[width=0.7\columnwidth]{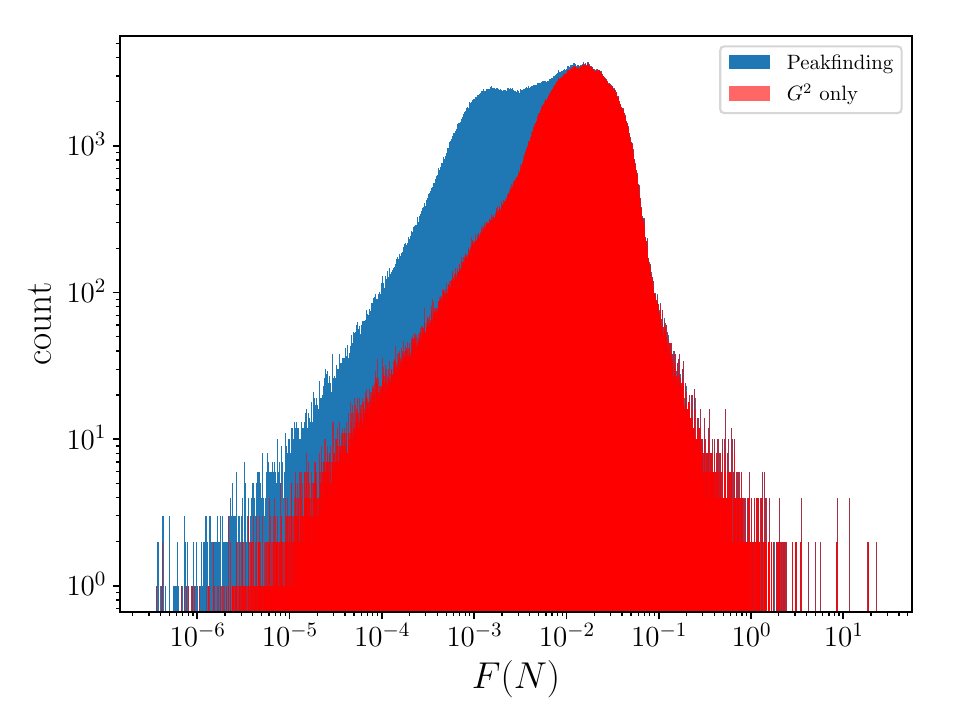}
    \caption{The histogram of forces using only the $G^2$ method (red) or the $G^2$ with peakfinding contact detection (blue).}
    \label{fig:histogram}
\end{figure}

This method allows us to detect forces that are an order of magnitude smaller than using the $G^2$ threshold alone. Using this method, we are able to account for weak contact forces that form a supporting network for the larger force chains. Figure~\ref{fig:contactnetwork} shows an example with the sample image provided by \citet{kollmer_photo-elastic_nodate}, where the red lines correspond to contacts detected using the $G^2$ threshold method ($G^2 = 0.5$) and the yellow lines show contacts detected using peak finding. We note that although we have increased the resolution of the forces (Fig.~\ref{fig:histogram}), we still systematically undercount some of the small contacts when the forces have an intensity peak smaller than the noise floor and are excluded (between particles 4 and 15 in Fig.~\ref{fig:contactnetwork}). 

\clearpage
\section{Statistical independence between branch and force vectors}

\begin{figure}[h]
    \centering
    \includegraphics[width=0.65\columnwidth]{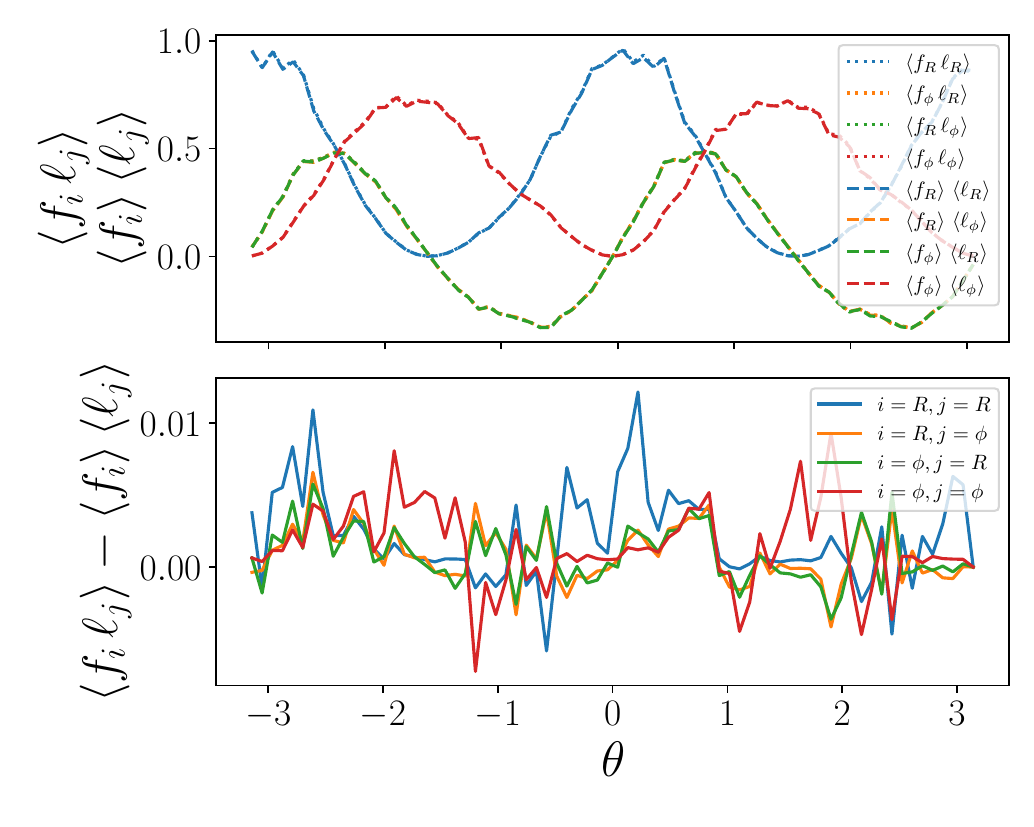}
    \caption{Top panel: comparing the directional statistical dependence of the force vector $\vec{f}$ and branch vector $\vec{\ell}$. We plot the averaged outer product and the outer product, averaged as a function of angle. The differences are small so we resolve this by plotting the residuals. Bottom panel: residuals showing that there is no angular dependence between the branch vector and force vector.}
    \label{fig:zeta}
\end{figure}

The branch vectors $\vec \ell$ and force vectors $\vec f$ are statistically independent, and thus:
\begin{equation}
\left< \ell_i f_j \right>(\theta) = \zeta \left< \ell_i\right>(\theta) \left< f_j\right>(\theta) 
\label{Eq_ksi}
\end{equation}

where $\zeta$ is a direction-independent scalar that is typically of order 1. We tested this assumption with our experimental data and found that there was no directional dependence between the two vectors and can use the right hand side of the equation with $\zeta = 1$. This is shown in Fig.~\ref{fig:zeta}, where we plot the left and right hand size of equation~\ref{Eq_ksi} in the top panel. The bottom panel shows the residuals between the right and left hand side of the equation, which, within noise, approximate zero indicating that $\zeta \approx 1$. These data are shown for the annular shear loading geometry, but is representative of all loadings tested in this study.

\section{Videos of loading evolving with strain}

To visualize the effect of increasing strain for each loading geometry we refer to the following videos: uniaxial compression~\cite{univideo}, isotropic compression~\cite{isovideo}, pure shear~\cite{pureshearvideo}, and annular shear~\cite{annulusvideo}. For each geometry, we show the photoelastic image along with the angular distributions of the contacts, average normal forces and average tangential forces. The strain steps correspond with the averaged strain windows indicated in the main text.